\documentclass{stanfest}
\newcommand{\msun}{\mbox{$M_{\odot}$}}
\newcommand{\Msun}{\mbox{$M_{\odot}$}}
\newcommand{\lsun}{\mbox{$L_{\odot}$}}

\newcommand{\teff}{\mbox{$T_{\rm eff}$}}

\newcommand{\vinf}{\mbox{$v_{\infty}$}}

\newcommand{\mdot}{\mbox{$\dot{M}$}}

\author{Jorick S.\  Vink}[AOP]
\affil[AOP]{Armagh Observatory and Planetarium, 
College Hill, BT61 9DG Armagh, Northern Ireland}
\title{Fifty years of CAK}

\begin{document}

\maketitle

\begin{abstract}
We present a new framework for massive star evolution that is no longer driven by Dutch or other mass-loss rate "recipes" but which take the physics of $\Gamma$ or $L/M$ dependent mass loss consistently into account. We first discuss the hot-star mass-loss kink and the transition mass loss rate between optically thin and thick winds, before discussing vertical stellar evolution, mass evaporation, and the maximum black hole (BH) mass. We end with a suggestion that a recently uncovered red supergiant (RSG) kink might be related to similar underlying $L/M$ physics as the hot-star kink.  
\end{abstract}

\section{Introduction}
\label{sec:intro}

Over the past five decades, the line-driven wind theories of \cite{LS70}
and \cite{CAK} or CAK have played a key role in our understanding of 
hot-star winds and massive star evolution. While these winds are intrinsically unstable 
the time-averaged behaviour in terms of the mass-loss rate (\mdot) and wind terminal velocity (\vinf) are rather 
well described by CAK theory \citep{OCR}.

However, CAK theory does rely on a number of assumptions that need to be improved upon (see \cite{vink22} for an overview). One such aspect is that the famous CAK $\alpha$ parameter that describes the line force from optically thin lines to
the total line force \citep{Puls08} is not constant in modern atmosphere codes \citep{sander20}. For optically thicker winds, the 
importance of multi-line scattering increases \citep{AL85,gayley95} and this process is not just relevant for classical Wolf-Rayet stars, but also for very massive stars (VMS) close to the Eddington limit.

The transition from optically thin O-star winds \citep{sundqvist19} to those of VMS \citep{vink11} is not only interesting from a wind physics perspective, but is also crucial for understanding stellar evolution and final products including the black hole mass function, as shown in the following.

\section{Mass loss kink}

\begin{figure}[!t]
\includegraphics[width=\textwidth]{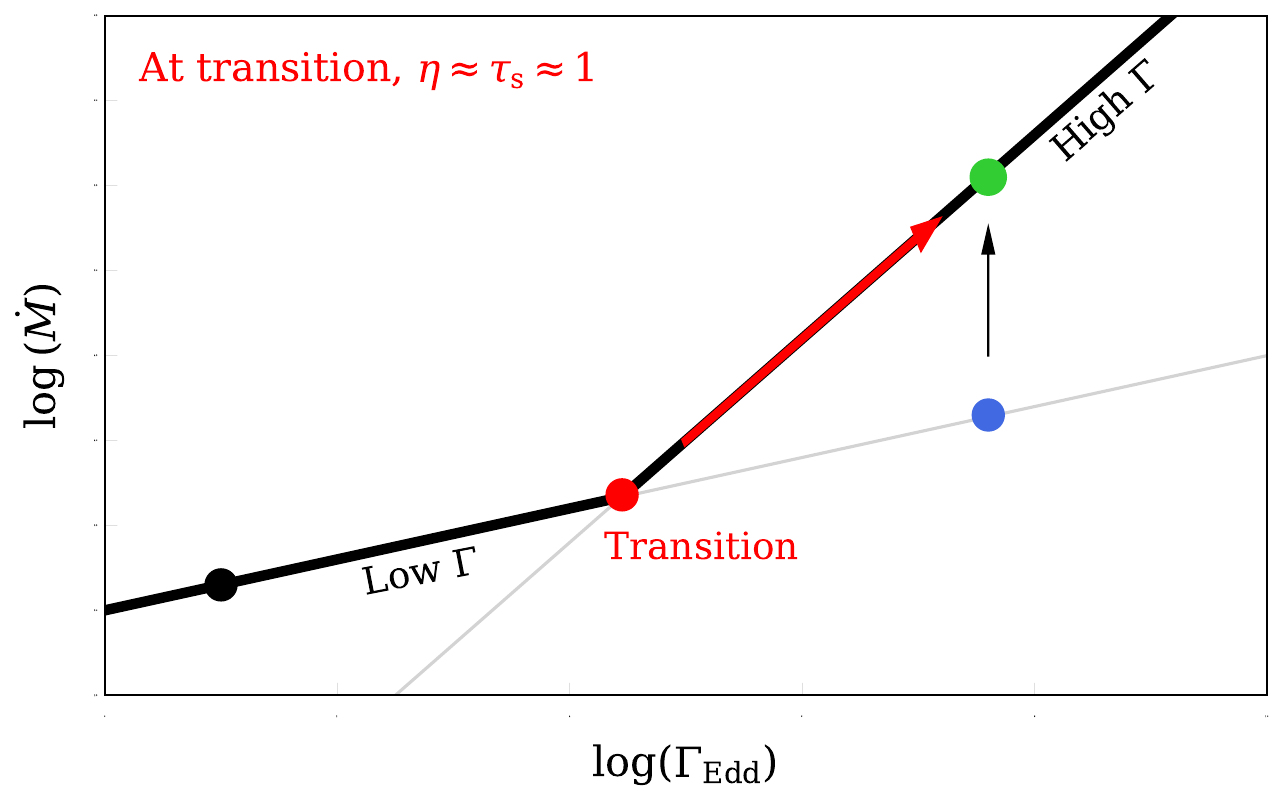}
\caption{A cartoon of the wind mass-loss rate versus the Eddington parameter $\Gamma$. The black dot represents stars in the traditional massive O-type regime that is relatively well-described by CAK theory, while the red dot provides the \cite{VG12} transition point. Above this red point, wind mass-loss rates need to be boosted from the blue to green dot, accounting for the much steeper \cite{vink11} slope at higher $\Gamma$.}
\label{fig:kink}
\end{figure}

Figure\,\ref{fig:kink} showcases the results of the theoretical \citep{vink11} and empirical \citep{best14} mass-loss kink at the transition from optically thin to optically thick winds for VMS up to 300\msun\ \citep{crowther10}.

For stars at the mass-loss transition point, \cite{VG12} showed that both the wind optical depth ($\tau$) and the wind efficiency number $(\eta)$, defined as:

\begin{equation}
    \eta = \frac{\mdot \vinf}{L/c}
\end{equation}
cross unity. In other words, $\eta = \tau = 1$ at the Of/WN spectra transition point\footnote{For detailed test computations, \cite{VG12} showed that there is a correction factor $f$ of 0.6 with only a small uncertainty of 0.2., enabling accurate mass-loss calibration.}. 

Interestingly, it was also shown that the traditional optically thin mass-loss relation \cite{vink00} exactly crosses this same mass-loss transition point, supporting the switch from the \cite{vink00} relation to the steeper high $\Gamma$ relation from \cite{vink11}, represented as:

\begin{equation}
    \log \dot{M} \propto 4.77 \log(L/L_{\odot}) -3.99 \log(M/M_{\odot})
\end{equation}

\begin{figure}[!t]
\includegraphics[width=\textwidth]{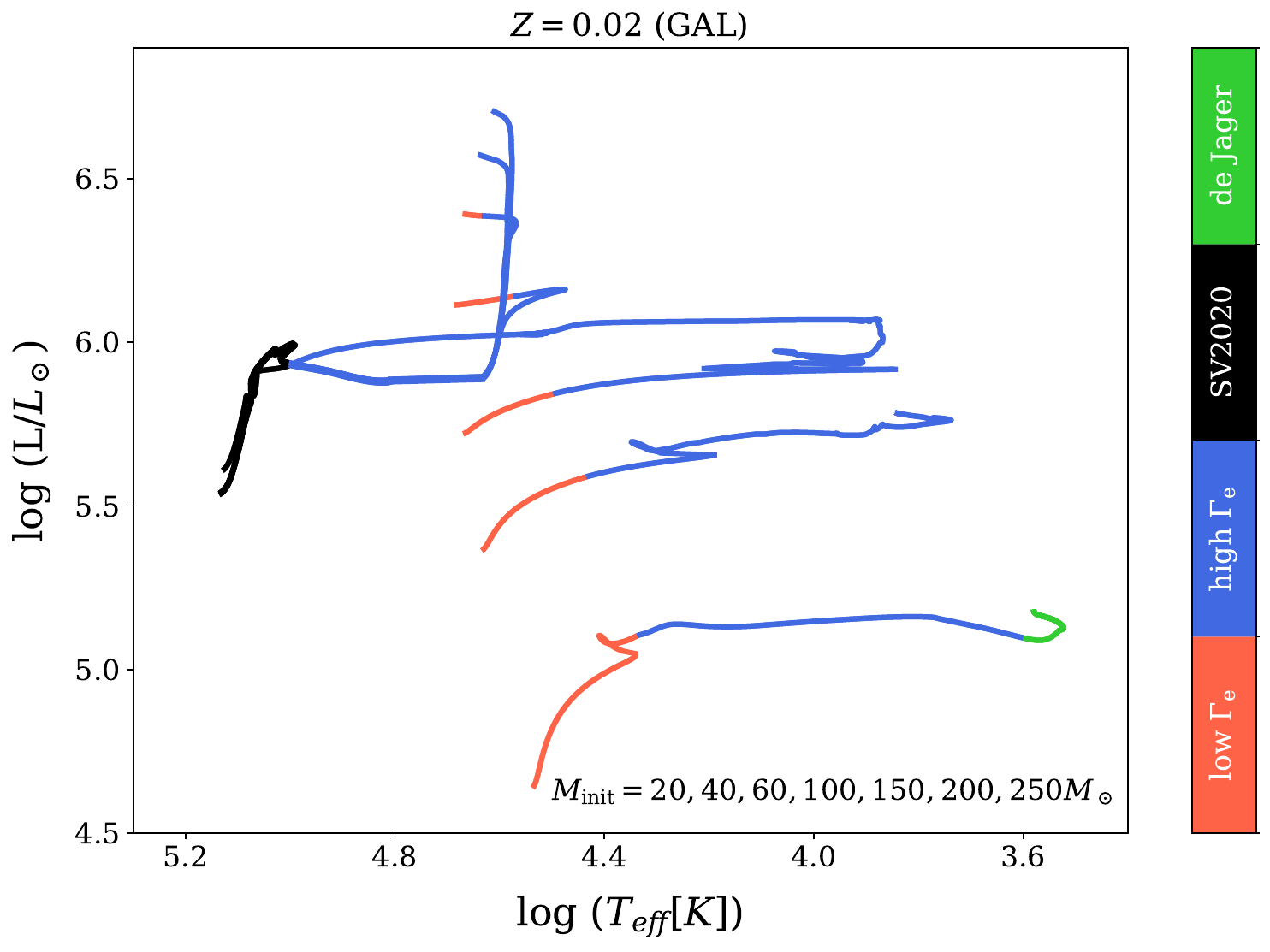}
\caption{The HR diagram for MESA models incorporates the mass-loss recipes shown on the right following the \cite{Sabh22} mass-loss implementation. Stars above the transition point exhibit vertical evolution, whereas lower-mass stars display the traditional horizontal evolution towards the red. The key point is that while the 20\msun\
model transitions into the red supergiant (RSG) phase, the more massive models do not. }.
\label{fig:HRD}
\end{figure}

For this reason, in Armagh we have recently implemented this mass-loss transition strategy into the MESA \citep{paxton13} stellar evolution code \citep{Sabh22,Sabh23}. Due to the strong mass-loss enhancement we obtain vertical HRD evolution at the highest masses (see Fig.\,\ref{fig:HRD}). Conversely, for stars of 100\msun\ and below stars undergo traditional horizontal evolution into the redder part of the stellar HR diagram. 

Note that vertical evolution may naturally explain the almost constant $\teff$ values of VMS in young stellar clusters such as R136 in the LMC, and NGC 3603 and the Arches cluster in the Galaxy. By contrast a mass-loss rate that is too low would result in strong envelope inflation effect, and produce VMSs that are too cool \citep{2012A&A...538A..40G}. 

Attempts to balance the effects of inflation (resulting in a redwards movement) and mass-loss stripping (leading in bluewards movement) would otherwise be almost impossible as one would not only need to play this balancing act for a single model, but for all stars with a wide range of masses and a variety of Eddington parameters. 
In other words, vertical evolution resulting from enhanced mass loss, appears to be the prime physical explanation to not only obtain the correct $\teff$ for one star but to get all stars at similar $\teff$ in the same young cluster environment containing VMSs. 

\section{Maximum Black hole mass}

\begin{figure}[!t]
\includegraphics[width=\textwidth]{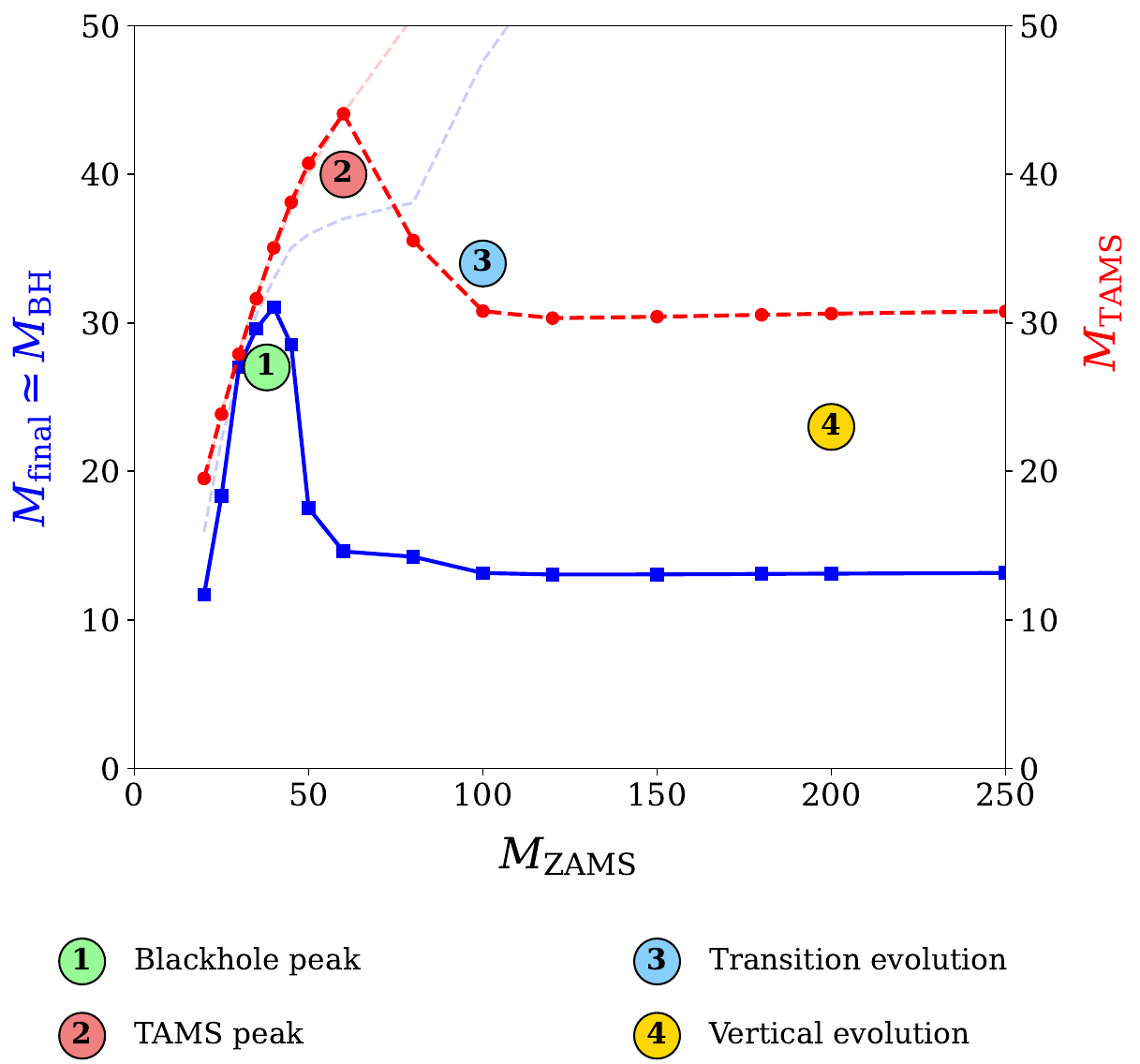}
\caption{The initial-final mass relation applies to initial ZAMS masses of 20, 25, 30, 35, 40, 45, 50, 60, 80, 100, 120, 150, 180, 200, and 250\msun.  The red and blue dashed lines represent the terminal main sequence (TAMS) and BH relationship assuming we only consider canonical low $\Gamma$ winds \cite{vink00}. Interestingly, the largest final BH masses are not found at the highest initial ZAMS masses; instead, they peak around $\simeq$40\msun, resulting in final BH masses of about 30\msun\ \citep{vink24}.}
\label{fig:MBH}
\end{figure}

The property of interest here is the final mass at the end of core helium burning, which we assume to correspond to the black hole (BH) mass \citep{Freyer12}. The initial-final mass results across the entire mass range are displayed in Fig.,\ref{fig:MBH}.

Interestingly, the largest BH mass (marked as "1" in Fig.\,\ref{fig:MBH}) is not found among the highest zero-age main sequence (ZAMS) masses, as one might expect. Instead, the maximum BH mass of 30\msun\ occurs within the modest ZAMS mass range of 35 to 45\msun. The 35\msun\ model loses only about 4\msun\	
  during core hydrogen burning, followed by an additional solar mass	
  during core helium burning.

The apparently counter-intuitive behavior of the constant BH mass tail ("4" in Fig.\,\ref{fig:MBH}) can be attributed to the inclusion of 
$\Gamma$-dependent VMS mass loss, which becomes significant for ZAMS masses above 80-100\msun.
This leads to the inflection point in the TAMS mass behavior noted at "Point 3" in Fig.\,\ref{fig:MBH}. For these high-mass stars, the mass-loss rates are so humongous that they not only strip away the hydrogen envelope but also reduce mass from the stellar core. Consequently, the BH masses from these stripped Wolf-Rayet stars are limited to about 10-15\msun\	
  \citep{Higg21} at solar metallicity.

  At lower $Z$ we not only expect the upper-mass limit of stars to be higher \citep{vink18}, but due the $Z_{\rm Fe}$ dependence of stellar winds during stellar evolution the maximum BH mass below pair instability to be as high as 93\msun \citep{Winch24}.

\section{Red supergiant kink}
\label{sec:rsg}

Red supergiants (RSGs) are essential for understanding the evolution of massive stars and their endpoints. However, uncertainties surrounding their mass-loss mechanisms have hindered the development of a comprehensive framework for massive star evolution. We analysed a recently identified empirical mass-loss "kink" feature, emphasizing its resemblance to radiation-driven wind models for hot stars and observations at the transition from optically thin-to-thick conditions, and we propose that a new mass-loss prescription for RSGs should accomodate the Eddington factor $\Gamma$ or at least an $L$ over current mass dependence $M_{\rm current}$. 

\begin{figure}[!t]
\includegraphics[width=\textwidth]{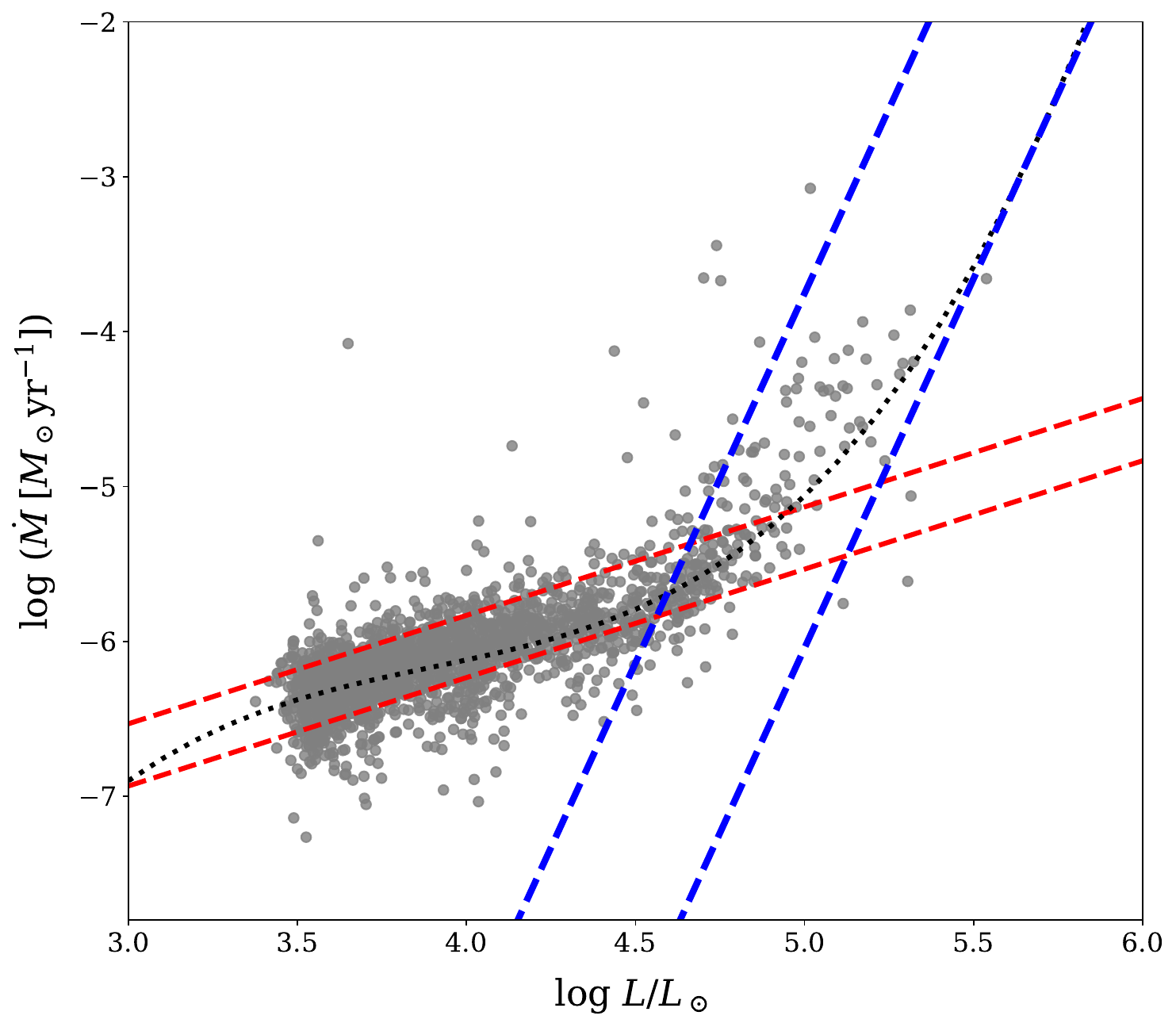}
\caption{Observational mass-loss rates from \citep{Yang23} show a kink at 
$\log(L/\lsun) = 4.6$. We utilize a 
$\Gamma$-dependent mass-loss relation informed by our understanding of radiation-driven winds, employing a shallow mass-loss relation below the kink and a steep one above it, as described by \citep{vink11}. The two blue lines represent current masses of 8\msun\	
  (upper line) and 30\msun.	
 (lower line). Additionally, a black dotted line indicates a third-degree polynomial fit in luminosity derived by \citep{Yang23}.}
\label{fig:RSG}
\end{figure}

Figure \ref{fig:RSG} presents observational data for RSGs in the Small Magellanic Cloud from \citep{Yang23}. While Yang et al. fitted their data with a third-order polynomial that solely depends on $L$, a key aspect of radiation-driven winds is their dependence on $\Gamma$, which is proportional to 
and inversely related to 
$M$. \cite{Beasor20} did propose a formula that shows a steep dependence on $L$ and an inverse relationship with 
$M$, but incorporating initial masses $M_{\rm init}$ (as derived from the overall properties of star clusters). We would instead expect that any luminous star that undergoes significant mass loss will increasingly exhibit stronger mass-loss rates, creating a positive feedback loop.
Incorporating both a steep dependence on luminosity, and a physically motivated inverse steep dependence on the current\footnote{While current masses cannot be readily obtained from observations this should not be an argument against implementation of an appropriate physical model. During the computation of a stellar track the mass evolution is simply followed by the model.} mass $M_{\rm current}$. 

The newly proposed RSG mass-loss prescription was implemented in the MESA stellar evolution code, and we found that our mass-loss framework correctly reproduces the Humphreys-Davidson limit (without requiring arbitrary adjustments) and simultaneously also resolves the RSG supernova problem. 

We suggest a universal trend observed in radiation-driven winds across the HR diagram, regardless of the specific source of opacity, is a crucial aspect of the evolution of the most massive stars.
 
\section{Summary}

Even after almost 5 decades the CAK theory still provides a useful framework for stationary winds of massive O stars \citep{OCR}. Above the \cite{VG12} mass-loss transition point where winds become optically thicker and multiple line scatterings become more dominant, the mass-loss rate versus Eddington $\Gamma$ relation becomes notably steeper. 

When such a mass-loss behaviour is included in MESA stellar evolution models this could lead to vertical rather than horizontal evolution in the stellar HR diagram. Mass evaporation effects result in a seemingly counter-intuitive result that it is not the most massive stars that produce the heaviest BHs. With our new mass-loss implementation, the maximum BH mass at Galactic metallicity is of order 30\Msun \citep{vink24}.

Finally, evolved supergiants stars, such as yellow and red supergiants (where $M$ and $L$ are decoupled) should probably also be subjected to $\Gamma$-dependent mass loss. As mass loss leads to a {\it smaller and smaller} current day mass -- for the same luminosity, this leads to a gradually {\it increasing} $\Gamma$ parameter, resulting in runaway mass-loss, enabling a superwind.

% ACKNOWLEDGMENTS:
%
\acknowledgements{We would like to thank Stan-the-Man for making CAK great again. We acknowledge MESA authors and developers
for their continued revisions and public accessibility of the code. 
We are supported
by STFC (Science and Technology Facilities Council) funding under grant numbers ST/V000233/1 and ST/Y001338/1 (PI Vink). 
This work made extensive use of NASA's Astrophysics Data System (ADS).}
%-----------------------------------------------------------------------------

%-----------------------------------------------------------------------------
% BIBLIOGRAPHY:
%
\bibliographystyle{stanfest_bibstyle}
\bibliography{stan_firstauthor}

\begin{thebibliography}{25}
\providecommand{\natexlab}[1]{#1}
\providecommand{\url}[1]{\texttt{#1}}
\providecommand{\urlprefix}{URL }
\providecommand{\eprint}[2][]{\url{#2}}

\bibitem[{{Abbott} \& {Lucy}(1985)}]{AL85}
{Abbott}, D.~C., {Lucy}, L.~B., \emph{\apj} \textbf{288}, 679 (1985)

\bibitem[{{Beasor} et~al.(2020)}]{Beasor20}
{Beasor}, E.~R., et~al., \emph{\mnras} \textbf{492}, 4, 5994 (2020)

\bibitem[{{Bestenlehner} et~al.(2014)}]{best14}
{Bestenlehner}, J.~M., et~al., \emph{\aap} \textbf{570}, A38 (2014)

\bibitem[{{Castor} et~al.(1975){Castor}, {Abbott}, \& {Klein}}]{CAK}
{Castor}, J.~I., {Abbott}, D.~C., {Klein}, R.~I., \emph{\apj} \textbf{195}, 157 (1975)

\bibitem[{{Crowther} et~al.(2010)}]{crowther10}
{Crowther}, P.~A., et~al., \emph{\mnras} \textbf{408}, 2, 731 (2010)

\bibitem[{{Fryer} et~al.(2012)}]{Freyer12}
{Fryer}, C.~L., et~al., \emph{\apj} \textbf{749}, 1, 91 (2012)

\bibitem[{{Gayley} et~al.(1995){Gayley}, {Owocki}, \& {Cranmer}}]{gayley95}
{Gayley}, K.~G., {Owocki}, S.~P., {Cranmer}, S.~R., \emph{\apj} \textbf{442}, 296 (1995)

\bibitem[{{Gr{\"a}fener} et~al.(2012){Gr{\"a}fener}, {Owocki}, \& {Vink}}]{2012A&A...538A..40G}
{Gr{\"a}fener}, G., {Owocki}, S.~P., {Vink}, J.~S., \emph{\aap} \textbf{538}, A40 (2012)

\bibitem[{{Higgins} et~al.(2021){Higgins}, {Sander}, {Vink}, \& {Hirschi}}]{Higg21}
{Higgins}, E.~R., {Sander}, A.~A.~C., {Vink}, J.~S., {Hirschi}, R., \emph{\mnras} \textbf{505}, 4, 4874 (2021)

\bibitem[{{Lucy} \& {Solomon}(1970)}]{LS70}
{Lucy}, L.~B., {Solomon}, P.~M., \emph{\apj} \textbf{159}, 879 (1970)

\bibitem[{{Owocki} et~al.(1988){Owocki}, {Castor}, \& {Rybicki}}]{OCR}
{Owocki}, S.~P., {Castor}, J.~I., {Rybicki}, G.~B., \emph{\apj} \textbf{335}, 914 (1988)

\bibitem[{{Paxton} et~al.(2013)}]{paxton13}
{Paxton}, B., et~al., \emph{\apjs} \textbf{208}, 1, 4 (2013)

\bibitem[{{Puls} et~al.(2008){Puls}, {Vink}, \& {Najarro}}]{Puls08}
{Puls}, J., {Vink}, J.~S., {Najarro}, F., \emph{\aapr} \textbf{16}, 3-4, 209 (2008)

\bibitem[{{Sabhahit} et~al.(2022){Sabhahit}, {Vink}, {Higgins}, \& {Sander}}]{Sabh22}
{Sabhahit}, G.~N., {Vink}, J.~S., {Higgins}, E.~R., {Sander}, A. A.~C., \emph{\mnras} \textbf{514}, 3, 3736 (2022)

\bibitem[{{Sabhahit} et~al.(2023){Sabhahit}, {Vink}, {Sander}, \& {Higgins}}]{Sabh23}
{Sabhahit}, G.~N., {Vink}, J.~S., {Sander}, A. A.~C., {Higgins}, E.~R., \emph{\mnras} \textbf{524}, 1, 1529 (2023)

\bibitem[{{Sander} et~al.(2020){Sander}, {Vink}, \& {Hamann}}]{sander20}
{Sander}, A. A.~C., {Vink}, J.~S., {Hamann}, W.~R., \emph{\mnras} \textbf{491}, 3, 4406 (2020)

\bibitem[{{Sundqvist} et~al.(2019){Sundqvist}, {Bj{\"o}rklund}, {Puls}, \& {Najarro}}]{sundqvist19}
{Sundqvist}, J.~O., {Bj{\"o}rklund}, R., {Puls}, J., {Najarro}, F., \emph{\aap} \textbf{632}, A126 (2019)

\bibitem[{{Vink}(2018)}]{vink18}
{Vink}, J.~S., \emph{\aap} \textbf{615}, A119 (2018)

\bibitem[{{Vink}(2022)}]{vink22}
{Vink}, J.~S., \emph{\araa} \textbf{60}, 203 (2022)

\bibitem[{{Vink} et~al.(2000){Vink}, {de Koter}, \& {Lamers}}]{vink00}
{Vink}, J.~S., {de Koter}, A., {Lamers}, H.~J.~G.~L.~M., \emph{\aap} \textbf{362}, 295 (2000)

\bibitem[{{Vink} \& {Gr{\"a}fener}(2012)}]{VG12}
{Vink}, J.~S., {Gr{\"a}fener}, G., \emph{\apjl} \textbf{751}, 2, L34 (2012)

\bibitem[{{Vink} et~al.(2024){Vink}, {Sabhahit}, \& {Higgins}}]{vink24}
{Vink}, J.~S., {Sabhahit}, G.~N., {Higgins}, E.~R., \emph{\aap} \textbf{688}, L10 (2024)

\bibitem[{{Vink} et~al.(2011)}]{vink11}
{Vink}, J.~S., et~al., \emph{\aap} \textbf{531}, A132 (2011)

\bibitem[{{Winch} et~al.(2024){Winch}, {Vink}, {Higgins}, \& {Sabhahitf}}]{Winch24}
{Winch}, E. R.~J., {Vink}, J.~S., {Higgins}, E.~R., {Sabhahitf}, G.~N., \emph{\mnras} \textbf{529}, 3, 2980 (2024)

\bibitem[{{Yang} et~al.(2023)}]{Yang23}
{Yang}, M., et~al., \emph{\aap} \textbf{676}, A84 (2023)

\end{thebibliography}
%-----------------------------------------------------------------------------

\end{document}